\documentstyle[12pt,epsf]{article}
\begin{document}

\catcode`@=11
\long\def\@caption#1[#2]#3{\par\addcontentsline{\csname
  ext@#1\endcsname}{#1}{\protect\numberline{\csname
  the#1\endcsname}{\ignorespaces #2}}\begingroup
    \small
    \@parboxrestore
    \@makecaption{\csname fnum@#1\endcsname}{\ignorespaces #3}\par
  \endgroup}
\catcode`@=12
\newcommand{\newc}{\newcommand}
\newc{\gsim}{\lower.7ex\hbox{$\;\stackrel{\textstyle>}{\sim}\;$}}
\newc{\lsim}{\lower.7ex\hbox{$\;\stackrel{\textstyle<}{\sim}\;$}}
\newc{\gev}{\,{\rm GeV}}
\newc{\mev}{\,{\rm MeV}}
\newc{\ev}{\,{\rm eV}}
\newc{\kev}{\,{\rm keV}}
\newc{\tev}{\,{\rm TeV}}
\def\tr{\mathop{\rm tr}}
\def\Tr{\mathop{\rm Tr}}
\def\Im{\mathop{\rm Im}}
\def\Re{\mathop{\rm Re}}
\def\bR{\mathop{\bf R}}
\def\bC{\mathop{\bf C}}
\def\lie{\mathop{\hbox{\it\$}}} 
\newc{\sw}{s_W}
\newc{\cw}{c_W}
\newc{\swsq}{s^2_W}
\newc{\swsqb}{s^2_W}
\newc{\cwsq}{c^2_W}
\newc{\cwsqb}{c^2_W}
\newc{\Qeff}{Q_{\rm eff}}
\newc{\mz}{m_Z}
\newc{\mpl}{M_{pl}}
\renewcommand{\phi}{\varphi}
\newc{\smu}{{\tilde\mu}}
\newc{\slep}{{\tilde\ell}}
\newc{\snu}{{\tilde\nu}}
\newc\order{{\cal O}}
\newc{\eps}{\epsilon}
\newc{\re}{\mbox{Re}\,}
\newc{\im}{\mbox{Im}\,}
\newc{\lunits}{\,\mbox{cm}^{-2}\mbox{s}^{-1}}
\newc{\etacp}{\eta_{\rm CP}}
%
%
\def\NPB#1#2#3{Nucl. Phys. {\bf B#1} (19#2) #3}
\def\PLB#1#2#3{Phys. Lett. {\bf B#1} (19#2) #3}
\def\PLBold#1#2#3{Phys. Lett. {\bf#1B} (19#2) #3}
\def\PRD#1#2#3{Phys. Rev. {\bf D#1} (19#2) #3}
\def\PRL#1#2#3{Phys. Rev. Lett. {\bf#1} (19#2) #3}
\def\PRT#1#2#3{Phys. Rep. {\bf#1} (19#2) #3}
\def\ARAA#1#2#3{Ann. Rev. Astron. Astrophys. {\bf#1} (19#2) #3}
\def\ARNP#1#2#3{Ann. Rev. Nucl. Part. Sci. {\bf#1} (19#2) #3}
\def\MPL#1#2#3{Mod. Phys. Lett. {\bf #1} (19#2) #3}
\def\ZPC#1#2#3{Zeit. f\"ur Physik {\bf C#1} (19#2) #3}
\def\APJ#1#2#3{Ap. J. {\bf #1} (19#2) #3}
\def\AP#1#2#3{{Ann. Phys. } {\bf #1} (19#2) #3}
\def\RMP#1#2#3{{Rev. Mod. Phys. } {\bf #1} (19#2) #3}
\def\CMP#1#2#3{{Comm. Math. Phys. } {\bf #1} (19#2) #3}
\relax
%
%
%
\def\beq{\begin{equation}}
\def\eeq{\end{equation}}
\def\bea{\begin{eqnarray}}
\def\eea{\end{eqnarray}}
%
%
%
\def\boxeqn#1{\vcenter{\vbox{\hrule\hbox{\vrule\kern3pt\vbox{\kern3pt
\hbox{${\displaystyle #1}$}\kern3pt}\kern3pt\vrule}\hrule}}}
%
%
\def\qed#1#2{\vcenter{\hrule \hbox{\vrule height#2in
\kern#1in \vrule} \hrule}}
\def\half{{\textstyle{1\over2}}} 
%
%
%
%
\newc{\ie}{{\it i.e.}}          \newc{\etal}{{\it et al.}}
\newc{\eg}{{\it e.g.}}          \newc{\etc}{{\it etc.}}
\newc{\cf}{{\it c.f.}}
%
%
%
%
\def\CAG{{\cal A/\cal G}}
\def\CA{{\cal A}} \def\CB{{\cal B}} \def\CC{{\cal C}} \def\CD{{\cal D}}
\def\CE{{\cal E}} \def\CF{{\cal F}} \def\CG{{\cal G}} \def\CH{{\cal H}}
\def\CI{{\cal I}} \def\CJ{{\cal J}} \def\CK{{\cal K}} \def\CL{{\cal L}}
\def\CM{{\cal M}} \def\CN{{\cal N}} \def\CO{{\cal O}} \def\CP{{\cal P}}
\def\CQ{{\cal Q}} \def\CR{{\cal R}} \def\CS{{\cal S}} \def\CT{{\cal T}}
\def\CU{{\cal U}} \def\CV{{\cal V}} \def\CW{{\cal W}} \def\CX{{\cal X}}
\def\CY{{\cal Y}} \def\CZ{{\cal Z}}
%
%
%
%
%
\def\grad#1{\,\nabla\!_{{#1}}\,}
\def\gradgrad#1#2{\,\nabla\!_{{#1}}\nabla\!_{{#2}}\,}
\def\partder#1#2{{\partial #1\over\partial #2}}
\def\secder#1#2#3{{\partial^2 #1\over\partial #2 \partial #3}}
%
%
%
%
%
\def\ltap{\ \raise.3ex\hbox{$<$\kern-.75em\lower1ex\hbox{$\sim$}}\ }
\def\gtap{\ \raise.3ex\hbox{$>$\kern-.75em\lower1ex\hbox{$\sim$}}\ }
\def\gl{\ \raise.5ex\hbox{$>$}\kern-.8em\lower.5ex\hbox{$<$}\ }
\def\roughly#1{\raise.3ex\hbox{$#1$\kern-.75em\lower1ex\hbox{$\sim$}}}
%
%
%
%
\def\slash#1{\rlap{$#1$}/} 
\def\dsl{\,\raise.15ex\hbox{/}\mkern-13.5mu D} 
\def\delsl{\raise.15ex\hbox{/}\kern-.57em\partial}
\def\Ksl{\hbox{/\kern-.6000em\rm K}}
\def\Asl{\hbox{/\kern-.6500em \rm A}}
\def\Dsl{\hbox{/\kern-.6000em\rm D}} 
\def\Qsl{\hbox{/\kern-.6000em\rm Q}}
\def\gradsl{\hbox{/\kern-.6500em$\nabla$}}
%
%
\let\al=\alpha
\let\be=\beta
\let\ga=\gamma
\let\Ga=\Gamma
\let\de=\delta
\let\De=\Delta
\let\ep=\varepsilon
\let\ze=\zeta
\let\ka=\kappa
\let\la=\lambda
\let\La=\Lambda
\let\del=\nabla
\let\si=\sigma
\let\Si=\Sigma
\let\th=\theta
\let\Up=\Upsilon
\let\om=\omega
\let\Om=\Omega
\def\ph{\varphi}
%
%
%
\newdimen\pmboffset
\pmboffset 0.022em
\def\oldpmb#1{\setbox0=\hbox{#1}%
 \copy0\kern-\wd0
 \kern\pmboffset\raise 1.732\pmboffset\copy0\kern-\wd0
 \kern\pmboffset\box0}
\def\pmb#1{\mathchoice{\oldpmb{$\displaystyle#1$}}{\oldpmb{$\textstyle#1$}}
      {\oldpmb{$\scriptstyle#1$}}{\oldpmb{$\scriptscriptstyle#1$}}}
%
%
%
%
%
\def\bar#1{\overline{#1}}
\def\vev#1{\left\langle #1 \right\rangle}
\def\bra#1{\left\langle #1\right|}
\def\ket#1{\left| #1\right\rangle}
\def\abs#1{\left| #1\right|}
\def\vector#1{{\vec{#1}}}
\def\inv{^{\raise.15ex\hbox{${\scriptscriptstyle -}$}\kern-.05em 1}}
\def\pr#1{#1^\prime}  
\def\lbar{{\lower.35ex\hbox{$\mathchar'26$}\mkern-10mu\lambda}} 
\def\e#1{{\rm e}^{^{\textstyle#1}}}
\def\ee#1{\times 10^{#1} }
\def\om#1#2{\omega^{#1}{}_{#2}}
\def\imp{~\Rightarrow}
\def\coker{\mathop{\rm coker}}
\let\p=\partial
\let\<=\langle
\let\>=\rangle
\let\ad=\dagger
\let\txt=\textstyle
\let\h=\hbox
\let\+=\uparrow
\let\-=\downarrow
\def\dot{\!\cdot\!}
\def\vfilll{\vskip 0pt plus 1filll}
\def\thefootnote{\fnsymbol{footnote}}
%

\begin{titlepage}
\begin{flushright}
{IASSNS-HEP-98/30\\
hep-ph/9804355\\
April 1998\\
}
\end{flushright}
\vskip 2cm
\begin{center}
{\large\bf CP Violation, Higgs Couplings, and Supersymmetry}
\vskip 1cm
{\normalsize
K.S.~Babu\footnote{Current address: Department of Physics, Oklahoma State
  University, Stillwater, OK 74078}, Christopher
Kolda\footnote{Current address: Lawrence Berkeley National
  Laboratory, MS 50A-5101, One Cyclotron Road, Berkeley, CA 94720}, 
John March-Russell\footnote{Alfred P.~Sloan
Foundation Fellow. Current address: Theory Division, CERN, CH-1211,
Geneva 23, Switzerland} and Frank Wilczek}\\
\vskip 0.5cm
{\it School of Natural Sciences\\
Institute for Advanced Study\\
Princeton, NJ~08540\\}
\end{center}
\vskip .5cm
\begin{abstract}
Supersymmetric extensions of the standard model generically contain
additional sources of CP violation.  We discuss how at one loop a
potentially large CP violating coupling of the lightest
Higgs, $h^0$, to leptons is induced in the minimal supersymmetric
standard model (MSSM).  The CP violating couplings of $h^0$
in extensions of the MSSM, such as the next-to-minimal
supersymmetric standard model (NMSSM) are also considered.  We
indicate how this CP violation
might be observed; in particular a polarization-dependent
production asymmetry, in the context of a muon collider,
provides a means to access this coupling cleanly.  In the MSSM,
existing limits on the
electric dipole moment (EDM) of the electron, coupled with
standard universality assumptions,  severly constrains any
such signal. Nevertheless, extensions 
of the MSSM, such as the NMSSM, allow CP-violating signals as
large as 100\%.

\end{abstract}
\end{titlepage}
\setcounter{footnote}{0}
\setcounter{page}{1}
\setcounter{section}{0}
\setcounter{subsection}{0}
\setcounter{subsubsection}{0}


\section{Introduction}

Supersymmetric (SUSY) extensions of
the Standard Model generically contain several additional
CP-violating
phases beyond the usual Cabbibo-Kobayashi-Maskawa phase.
Elucidating their magnitude and structure is important if we
are properly to understand the origin of CP violation,
or the closely related question of
the origin of the cosmic matter-antimatter asymmetry.
In this paper we will be concerned with new sources of CP violation
in the Higgs sector of SUSY models.

Despite the large number of
new phases in the model as a whole,
it is
well known that in the minimal SUSY extension of the standard
model (MSSM) the tree level Higgs potential contains just one complex
parameter, the term $B \mu H_u H_d$.
Even this
phase can be removed by redefinitions of the Higgs fields.  Then
the vacuum expectation values of the Higgs scalars will be real, with
no mixing between the physical scalar and pseudoscalar Higgs fields.

At one loop the situation is different.
In Section~2 we will demonstrate that sizable Higgs sector
CP violation can be induced at one-loop, even within the MSSM, and
especially so at large $\tan\beta$.  One place where CP-violating effects
can manifest themselves is in the couplings of the lightest neutral
Higgs boson to Standard Model fermions.  In fact, we
will see that these are in principle accessible at a suggested muon collider
operating on the Higgs resonance, at least in some regions
of the SUSY parameter
space.  Alternatively, in variations of the MSSM with extended Higgs
sectors (such as the so-called NMSSM, defined below)
the tree-level Higgs potential contains
irremovable CP violating phases.  Then  substantial
CP violation is possible even for small  $\tan\beta$.

The possible magnitude of CP violation in the Higgs
sector is severely constrained by experimental limits on
electric dipole moments (EDMs) of fermions.
This is because the same diagrams that contribute to the
CP-violating Higgs couplings
also contribute to the EDMs of fermions.  As we discuss in Section~3, current
bounds on the $\mu$EDM do little to constrain our scenario, but
within the MSSM, bounds on the
$e$EDM are highly constraining given minimal
theoretical prejudices.
In Section~3 we therefore also discuss
how observation of CP violation, or lack thereof, in the Higgs-lepton coupling
fits into the broader picture painted by flavor-changing and CP-violation
constraints on SUSY extensions of the Standard Model.  We emphasize that
CP violation in the Higgs-fermion couplings
is a way of discriminating the MSSM from more elaborate extensions, such as
the NMSSM.  Our conclusions are summarized in Section~4.

Though most of our discussion is phrased in terms of the coupling
of the lightest neutral Higgs, $h^0$, to charged leptons, much of our
analysis applies more generally with only slight alterations
to Higgs-quark couplings.  There are also potentially interesting
effects of CP violation in the heavy Higgs sector, which are under study.

\section{CP Violation in the Higgs Sector}

\subsection{The MSSM}

Let us first discuss the situation in the minimal supersymemtric
extension of the standard model (MSSM).

In the absence of SUSY breaking, the
charged leptons couple to the Higgs field $H_d^0$, but not $H_u^0$.
After SUSY breaking, a coupling of the leptons to $H_u^0$ is generated
at one loop, and in general it will not be real, due to phases in the
soft-breaking parameters and the $\mu$ term.  The resulting coupling
of the lepton to the Higgs fields is of the form
\beq
{\cal L}=a\,\bar\ell_L\ell_R H_d^0+b\,\bar\ell_L\ell_R H_u^{0*}+h.c.
\label{eq:Leff}
\eeq
By redefining lepton fields while keeping the vacuum expectation values
of the Higgs fields real,
we may choose $a$ real but must then
allow $b$ to be complex.
The resulting lepton mass term is then:\footnote{In general, because
of SU(2)${}_L$-breaking, one should write
one set of terms for the lepton mass generation and another set for their
Yukawa couplings to the Higgses. However, if the electroweak symmetry-breaking
effects are small ($m_Z$ smaller than $M_{\rm SUSY}$), both the mass term and 
the Higgs coupling will to a good approximation arise from the terms in
Eq.~(\ref{eq:Leff}).}
\bea
{\cal L}_{\rm mass}&=&\bar\ell_L(av_d+bv_u)\ell_R + h.c. \nonumber\\
&=&\re(av_d+bv_u)\bar\ell\ell+i\im(av_d+bv_u)\bar\ell\gamma^5\ell~,
\eea
where $v_{u,d}=\vev{H^0_{u,d}}$.
In the Standard Model with only one Higgs doublet, no physical
CP violation can arise from the Higgs coupling,
because the same rotations that make the lepton masses real
also make their couplings to the neutral Higgs particle real.
But in this two-doublet extension the fermion masses do not correspond
directly to the couplings to the physical Higgs states.
Specifically,
we can write
\bea
\re\left(H_u^0\right)
&=&\frac{1}{\sqrt2}\left(\cos\alpha\, h^0+\sin\alpha\, H^0\right)
\nonumber \\
\re\left(H_d^0\right)
&=&\frac{1}{\sqrt2}\left(-\sin\alpha\, h^0+\cos\alpha\, H^0\right)
\eea
where $m_{h^0}<m_{H^0}$.

Here, for the sake of simplicity,
we have made the good approximation of omitting
additional ``pseudoscalar'' components on the right hand side.
In principle
one loop contributions to the Higgs effective potential can
spontaneously break CP~\cite{maekawa}, 
and/or communicate explicit CP-violation in the soft masses and $\mu$-term
to the Higgs sector. In either case, a small
phase in the Higgs vacuum expectation values is induced which
leads to scalar-pseudoscalar mixing, but this effect is $\lsim 1 \%$,
too small to affect our conclusions.  A relative phase between
$\left \langle H_u \right \rangle $ and $\left\langle H_d \right \rangle $
arises in the MSSM when a Higgs quartic coupling term
$(H_u H_d)^2$ is induced, but since supersymmetry is broken softly,
such quartic terms arise from finite box graphs, which
lead to a very small coefficient.

Since the transformations that
eliminate phases in the lepton and quark masses do not eliminate
phases from their couplings to $h^0$ and $H^0$, there is a residual
violation of CP.
In order to extract the CP violating portion of the $h^0\bar\ell\ell$
coupling, we must look for some mismatch between its phase and the
phase of the $\bar\ell\ell$ mass term. It is convenient to write the
Higgs coupling and mass terms in the forms
$(1/\sqrt{2})|-a\sin\alpha + b\cos\alpha|h^0\bar\ell
e^{i\gamma^5\phi}\ell$ and
$|av_d+bv_u|\bar\ell e^{i\gamma^5\delta}\ell$
respectively, where
\bea
\tan\phi&=&\frac{\im(-a\sin\alpha+b\cos\alpha)}
{\re(-a\sin\alpha+b\cos\alpha)} \\
\tan\delta&=&\frac{\im(a\cos\beta+b\sin\beta)}{\re(a\cos\beta+b\sin\beta)}.
\eea
When the mass term is made real through a chiral rotation of the $\ell$-field,
the coupling to the light Higgs becomes
\beq
\CL_{h\bar\ell\ell}=\frac{1}{\sqrt{2}}\left|-a\sin\alpha+b\cos\alpha\right|
h^0\bar\ell e^{i(\phi-\delta)\gamma^5}\ell.
\eeq
The observable CP-violating phase is then $\phi-\delta$. It is clear that in
the limit $\alpha\to\beta-\frac{\pi}{2}$, or equivalently
$\cot\alpha\to-\tan\beta$, the phase $\phi-\delta$ disappears
(recall $\tan\beta \equiv v_u/v_d$).  This is the
well-known Higgs decoupling limit of the MSSM in which the second doublet
becomes much heavier than the weak scale ($m_A\gg\mz$) so that the low-energy
Higgs sector closely approximates that of the SM.  The Higgs mixing
angle aligns itself with that of the fermion mass terms so that the fermions
effectively couple to only one scalar Higgs field, allowing all phases to be
removed.

\begin{figure}
\centering
\centerline{
\epsfysize=1.75in
\hfill\epsfbox{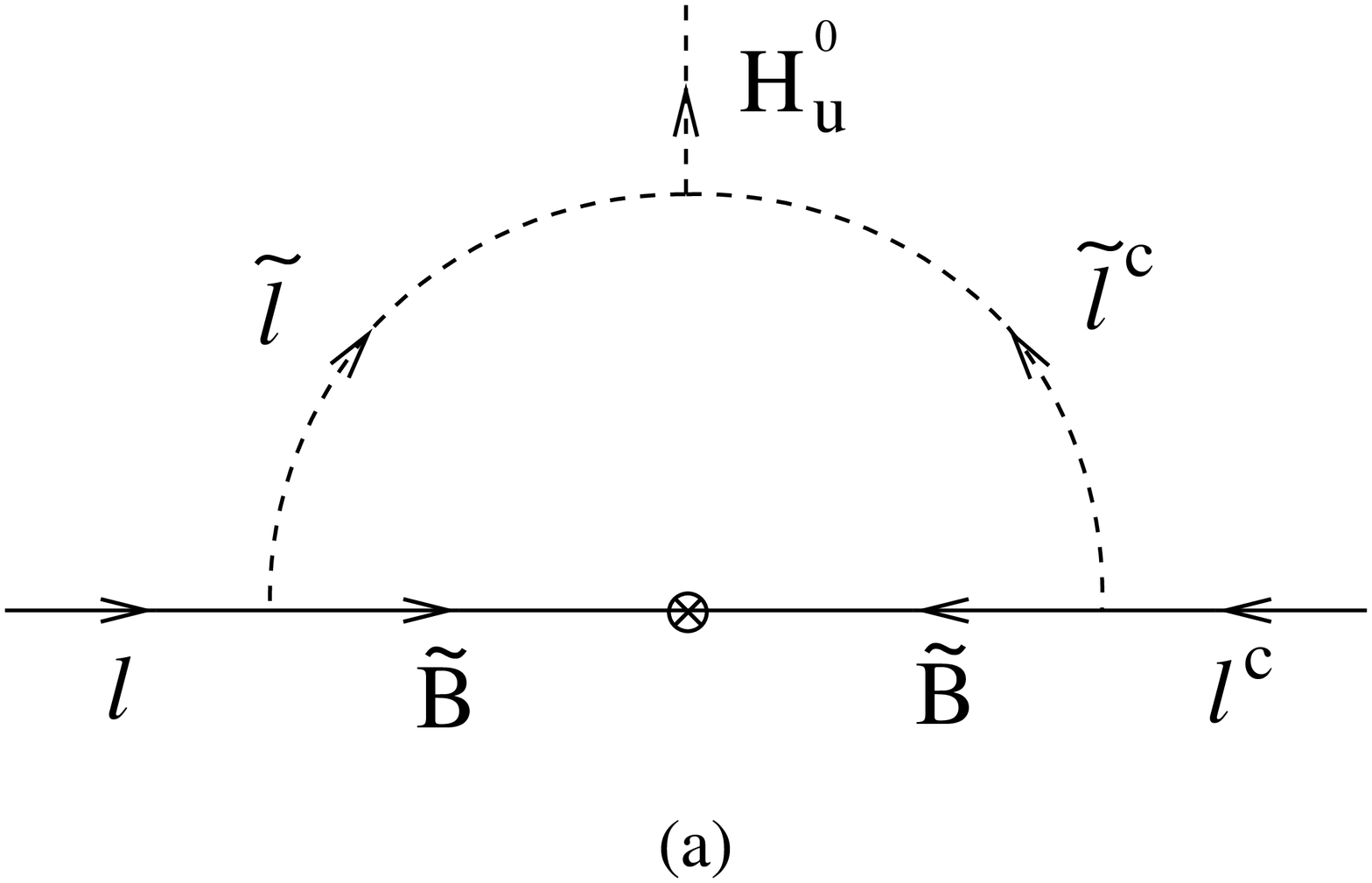}\hfill
\epsfysize=1.75in
\hfill\epsfbox{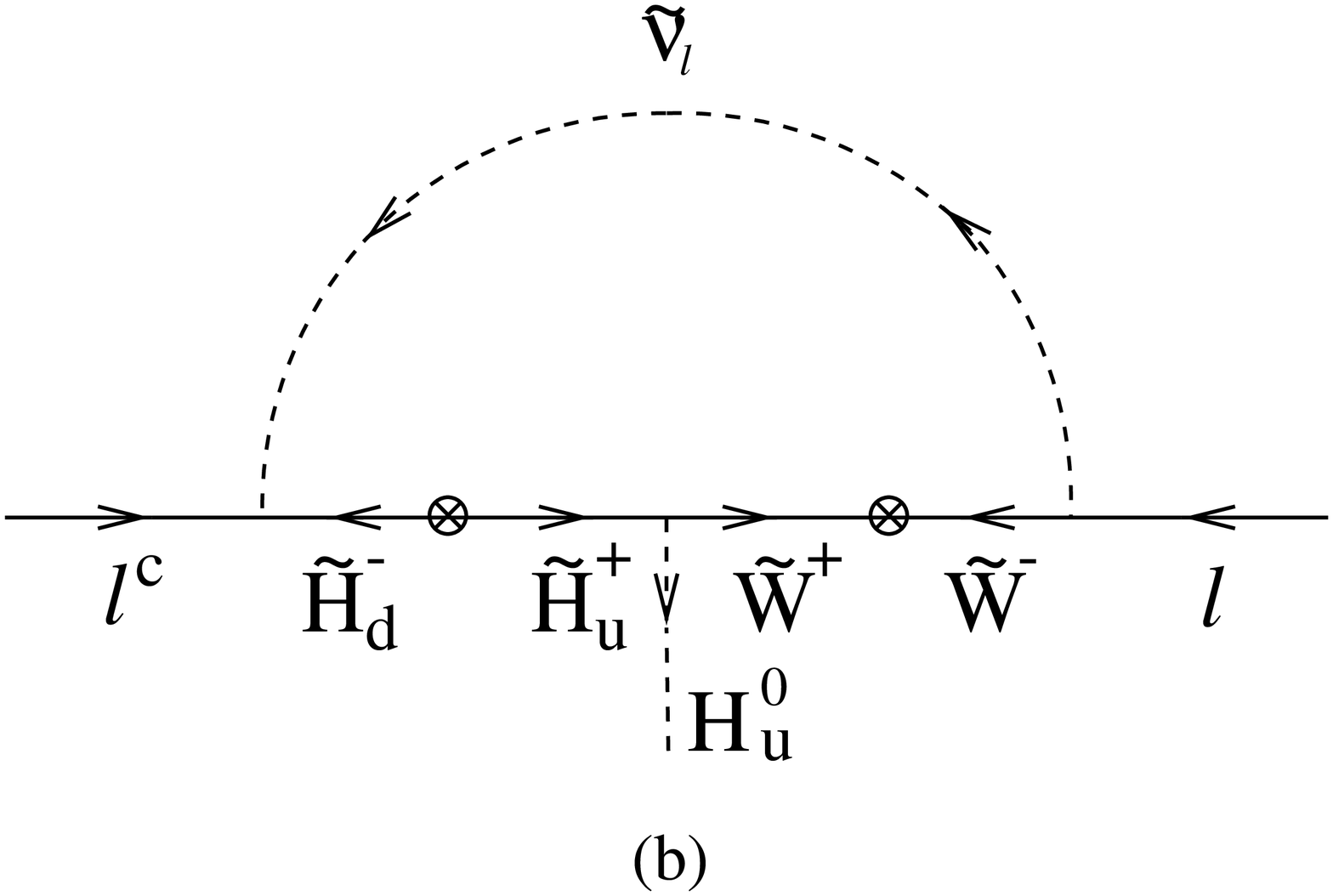}\hfill
}
\caption{Representative diagrams which contribute to the $\bar\ell_L\ell_R
H_u^{0*}+h.c.$ coupling.}
\label{fig:diagrams}
\end{figure}

We are interested in the CP violation in the Higgs sector
arising from the one loop induced parameter $b$.
Representatives of the two basic classes of
diagrams that contribute to the $\bar\ell\ell H_u^{0*}$ amplitudes are
shown in Fig.~\ref{fig:diagrams}. These are in the same class of
diagrams whose real parts have been studied both in certain fermion mass
generation scenarios~\cite{banks} and have been found to significantly
shift the $b$-quark mass at large $\tan\beta$~\cite{hrs}; here we will 
confine ourselves to the leptonic sector of the theory and study the 
imaginary parts of the diagrams.
We will only work to leading order in the
slepton, neutralino and chargino mixing, which is to say $\mz\ll\mu, M_2,
m_{\smu}$, and in the limit of $\tan\beta\gg1$.
In these limits, all of the contributing diagrams are linearly
dependent on $\mu^*/\cos\beta$, though for different
reasons. (Contributions proportional to soft trilinear $A$-terms do
not receive this $1/\cos\beta$ enhancement and thus their
contributions to Higgs sector CP violation are unobservably
small in generic scenarios.)  This is also the
approximation in which our effective Lagrangian is adequately described by
Eq.~(\ref{eq:Leff}); that is, we ignore terms in $\CL$ with multiple
insertions of $H_{u,d}^{0\dagger} H^0_{u,d}$ suppressed by the
SUSY-breaking scale.
This approximation is sufficient for our purposes since in
the region where the CP violating Higgs couplings of the MSSM are significant,
the
corrections due to exact diagonalization of the various mass matrices are
small.

The contribution to the Yukawa coupling from
diagram~\ref{fig:diagrams}(a) 
contains a factor of $\mu^*/\cos\beta$ coming from
the left-right mixing
of the sleptons\footnote{In order to set our sign convention for $\mu$,
we take $W=\mu(H_d^-H_u^+ - H_d^0 H_u^0)$.}:
\beq
A_1^\ell=\frac{3\alpha_1}{20\pi}y_\ell\mu^*M_1 f\left(M_1^2,m_\slep^2,
m_{\slep^c}^2\right)
\label{eq:A1}
\eeq
where $y_\ell$ is the lepton Yukawa coupling, $M_1$ is the U(1) gaugino (\ie,
bino) mass and
\beq
f\left(m_1^2, m_2^2, m_3^2\right)\equiv\frac{1}{m_3^2}\left[\frac{x\ln x}{1-x}
-\frac{y\ln y}{1-y}\right]
\frac{1}{x-y}
\eeq
with $x=m_1^2/m_3^2$ and $y=m_2^2/m_3^2$. Diagram~\ref{fig:diagrams}(b) 
picks up a $1/\cos\beta$
from the Yukawa coupling of the external muon to the $\tilde H_d$ higgsino and
a $\mu^*$ from the mixing of the $\tilde H_d$ with $\tilde H_u$ on the internal
line:
\beq
A_2^\ell=\frac{\alpha_2}{8\pi}y_\ell\mu^*M_2\left[f\left(\mu^2,
m^2_\slep, M_2^2\right) + 2f\left(\mu^2,m_{\snu}^2,M_2^2\right)\right]
\label{eq:A2}
\eeq
where $M_2$ is the SU(2) gaugino (\ie, wino) mass.  The contributions from both
the charged and neutral gaugino/higgsino loops are included in $A_2^\ell$. Finally,
there are also contributions analogous to those of 
diagram~\ref{fig:diagrams}(b) but including
only the (neutral) bino states in the loops.  They contribute to the amplitude
\beq
A_3^\ell=-\frac{3\alpha_1}{40\pi}y_\ell\mu^*M_1\left[f
\left(\mu^2, m^2_\slep, M_1^2\right) - 2f\left(\mu^2, m_{\slep^c}^2,
M_1^2\right)\right].
\label{eq:A3}
\eeq
Note that the function $\mu^* M_2 f(\mu^2, m_\slep^2, M_2^2)$
in Eq.~(\ref{eq:A2}) (and similarly in Eq.~(\ref{eq:A3})) has a maximum
value of 1.  In contrast, the function $\mu^* M_1 f(M_1^2, m^2_\slep,
m^2_{\slep^c})$ of Eq.~(\ref{eq:A1})
has a maximum value of $\mu^*/M_1$ which can be significantly larger than 1.

The parameter $b$ in the effective Lagrangian is then simply the sum of the
$A_i^\ell$'s. Numerically, $|b\tan\beta|\ll a$ (since $b$ is loop-suppressed)
so that we can approximate
\bea
\etacp^\ell\equiv
\tan(\phi-\delta)&\simeq&-\frac{\im(b)}{a}\left(\cot\alpha
+\tan\beta\right) \\
&\simeq&-2\frac{\im(\sum\!A_i^\ell)}{y_\ell}\,\frac{\min(m_A^2,\mz^2)}
{m_A^2+\mz^2}
\label{eq:tan}
\eea
where $\etacp^\ell$ will be used henceforth to parameterize the amount of CP
violation in the Higgs-lepton couplings.
To get this last equation, we have used the well-known relation~\cite{HHG} of
the MSSM, $\sin2\alpha\simeq-(m_A^2+\mz^2)/|m_A^2-\mz^2| \sin2\beta$ 
for large $\tan\beta$. Note that
Eq.~(\ref{eq:tan}) demonstrates explicitly the suppression of the CP violation
in the Higgs decoupling (large $m_A$) limit.

For the case of the MSSM, lower $\tan\beta$
means proportionally smaller $\etacp^\ell$. Therefore we will restrict our
attention to the large $\tan\beta$ regime, in which $\etacp^\ell$ is maximized,
when discussing the MSSM.
Of course, even if the underlying CP violating phase $\arg(\mu)$ is $\CO(1)$,
the effects in the Higgs-lepton couplings will always be suppressed by a loop
factor.  Thus even at $\tan\beta\sim50$ one does not expect more than a 10\%
effect in the Higgs couplings, \ie, $\etacp^\ell\lsim0.1$.

On the other hand, for the down-type quarks in the MSSM, the one-loop
induced CP violation in the Higgs-quark couplings can be substantially
bigger (of order 100\%) owing to the $\tan\!\beta$-enhanced contribution
from the gluino.

\begin{figure}
\centering
\epsfysize=3in
\hspace*{0in}
\epsffile{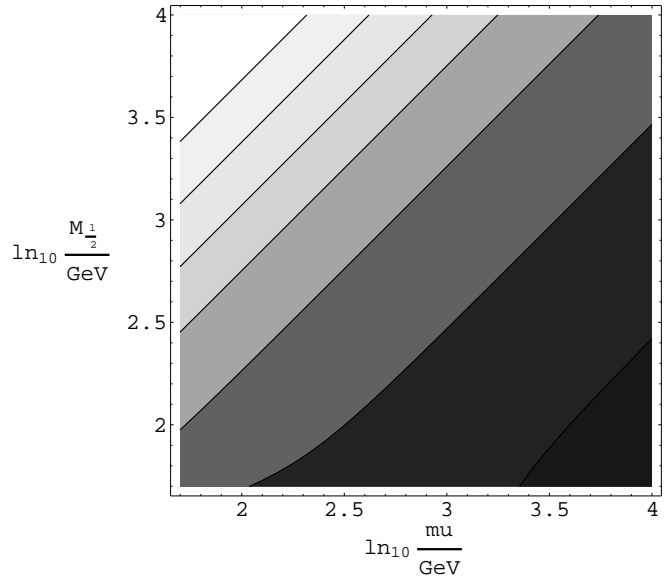}
\caption{Plot of $\etacp^\mu$ in the MSSM as a function of $|\mu|$ and the
gaugino mass parameter $M_{1/2}$, both varying on a log scale
from $50\gev$ to $10\tev$.  We have chosen $\tan\beta=50$,
$\arg(\mu)=\pi/4$ and
the smuon soft mass parameter $M_0 = 2M_{1/2}$.  The shaded regions 
correspond to
$\etacp^\mu>10\%$, 5\%, 2.5\%, 1\%, 0.5\%, 0.25\%, and 0.125\% going
from darkest to lightest.}
\label{fig:eta}
\end{figure}

In Figure~\ref{fig:eta}\ we have shown a contour plot
of the value of $\etacp^\mu$ for
$\tan\beta =50$ and $\arg(\mu)=\pi/4$, with $|\mu|$ along the $x$-axis.
For simplicity we parameterize the masses of the gauginos along the
$y$-axis using the usual
supergravity-type parameter, $M_{1/2}$, where
gaugino mass unification is
assumed, \ie, $M_i = \alpha_i M_{1/2}/\alpha_{\rm unif}$.
We also assume scalar mass unification, defining a ``common'' scalar
mass $M_0$; however this last assumption is only used to define
$m^2_{\tilde\ell_L}=M_0^2+0.5M_{1/2}^2$ and $m^2_{\tilde\ell_R}=
M_0^2+0.15M_{1/2}^2$
which are the formulas which follow from a renormalization-group
analysis in supergravity models.
For this figure we make the further illustrative choice that $M_0=2M_{1/2}$,
though the qualitative features of the figures are independent of this,
or any other, simplification.
Both $|\mu|$ and $M_{1/2}$ are allowed to vary from $50\gev$ to $10\tev$
logarithmically.

Note that the biggest effects occur
when the SUSY masses are large compared to the
weak scale, with both Higgs doublets remaining light. This is not necessarily
an unnatural scenario.  Indeed,
within supergravity-mediated models of SUSY-breaking,
one expects the Higgs potential at large $\tan\beta$ to be extremely flat.
In this case the second derivative along the imaginary direction (\ie,
$m_A^2$) will be small, ensuring light Higgs doublets.

\subsection{The NMSSM} \label{sec:nmssm}

The next-to-minimal supersymmetric standard model (NMSSM) is the
simplest extension of the MSSM.  In this model, the $\mu H_d H_u$
term is replaced by the superpotential interaction
$\Delta W = \lambda NH_d H_u +\frac{k}{3} N^3$, where $N$ is a gauge
singlet \cite{fayet}.  This has the advantage that the explicitly dimensionful
parameter $\mu$ of the MSSM is replaced by $\lambda \vev{N}$, with $\lambda$
dimensionless.
For our present purposes,
the most important difference from the MSSM is that now there is an
irremovable phase in the Higgs sector, which can generate large CP
violation even if all the soft supersymmetry breaking
parameters are approximately real.  Explicitly, the Higgs potential now has
three terms with non-trivial phase structure: $V \supset
\lambda k^* H_u H_d N^{* 2} + A_\lambda \lambda H_u H_d N
+ A_k k N^3 + h.c.$\  Even if the soft terms
$A_\lambda \lambda $ and $A_k k$ are
real, the phase of $\lambda k^*$ cannot be removed.  This generally leads
to complex vev's for $H_u$, $H_d$ and $N$~\cite{haba}.
The physical Higgs particles will then be admixtures of scalars and
pseudoscalars.

Thus in the NMSSM there are, already at tree-level, CP-violating couplings of
the mass eigenstate Higgs particles to fermions.  This follows simply from
\bea
{\cal L}&=&a\,\bar\ell_L\ell_R H_d^0 + h.c. \nonumber\\
&=&\frac{a}{\sqrt{2}}\left(O_{11}\bar\ell\ell h^0 -
iO_{21}\bar\ell\gamma^5\ell h^0\right)~,
\eea
where $O_{ij}$ is the matrix diagonalizing the Higgs sector mass eigenstates
in the basis $(\re H_d^0, \im H_d^0, \re H_u^0, ...)^T = O (h^0,...)^T$.
We can again define $\etacp^\ell$
for the NMSSM (or any larger extension of the MSSM) as the amount of CP
violation present in the Higgs-lepton couplings:
$\etacp^\ell \equiv - O_{21}/O_{11}$, assuming $\etacp^\ell\ll1$. 
(The above arguments follow equally well for
Higgs couplings to quarks in the NMSSM.)

Now, however, in contrast to the MSSM, $\etacp$ has
no strong dependence on $\tan\beta$, and no loop suppression, so that there
can be large CP-violating effects
over a wide range of $\tan\beta$.

\subsection{Collider Searches} \label{sec:collider}

In order to measure directly the amount of CP violation in the Higgs-lepton
couplings, one will undoubtedly require a very large number of well-tagged
Higgs bosons. To our knowledge, the most promising scheme for producing
such a large sample is to operate a muon collider on the Higgs
resonance\footnote{For a related discussion on CP violation in two photon
coupling of the Higgs, see Ref. \cite{gunion}.}.
There are in principle several CP-violating observables which are accessible in
a muon facility~\cite{soni}. The
most straightforward analysis is for the left-right polarization
production asymmetry:
\beq
\CA=\frac{\sigma(\mu^+_L\mu^-_L\to h^0\to \bar bb)-
\sigma(\mu^+_R\mu^-_R\to h^0\to\bar bb)} {\sigma(\mu^+_L\mu^-_L\to h^0\to\bar
bb)+
\sigma(\mu^+_R\mu^-_R\to h^0\to\bar bb)}\equiv \frac{\sigma_{LL}-
\sigma_{RR}}{\sigma_{LL}+\sigma_{RR}}
\eeq
which is zero in the absence of CP violation.
It is simple to show that
\beq
\CA\simeq\frac{4\etacp^\mu P}{1+P^2}.
\label{eq:ALR}
\eeq
for beam polarizations $P$, assuming the same polarization for both beams
and $\etacp^\mu \ll 1$.
This is to be compared to a background from
$\mu^+\mu^-\to(\gamma,Z)\to\bar bb$,
which has a cross-section of the same order as that of the Higgs-mediated
process, but suppressed by $(1-P^2)$. A simple estimate can be made
of the integrated luminosity, $\int\!\CL$, that will be necessary in order to
make a $3\sigma$ discovery of non-zero $\CA$
(without considering losses due to acceptances and efficiencies):
$\int\!\CL=(3/\CA)^2(\sigma_S+\sigma_B)/\sigma_S^2$
where $\sigma_{S(B)}$ is the signal (background) cross-section
$\sigma_{LL}+\sigma_{RR}$.  

\begin{figure}[th]
\centering
\epsfxsize=3.5in
\hspace*{0in}
\epsffile{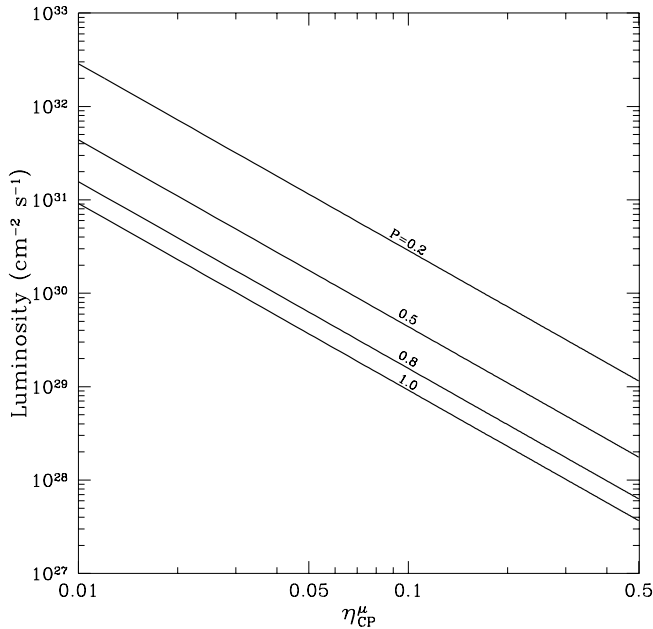}
\caption{Luminosity needed for a $3\sigma$ measurement of CP violation in
the polarized production asymmetry $\mu^+_{L,R}\mu^-_{L,R}\to h^0
\to\bar bb$ in one year, as a
function of the amount of CP violation in the Higgs-muon coupling 
($\etacp^\mu$).
The four lines correspond (in descending order) to beam polarizations 
$P=0.2$, 0.5, 0.8 and 1.0.}
\label{fig:ALR}
\end{figure}

In Fig.~\ref{fig:ALR} we plot the luminosity needed for a $3\sigma$ measurement
of non-zero $\CA$ in one year ($10^7\,$s) against our CP-violating
parameter $\etacp^\mu$. 
(The figure assumes $m_{h^0}=100\gev$, but varying $m_{h^0}$
changes the figure little.)
The plotted lines represent the limit for different beam polarizations:
from top to bottom, $P=0.2, 0.5, 0.8$ and $1.0$. For current collider design
parameters of $\CL=5\times10^{30}\lunits$ and ``natural'' beam polarization
($P=0.2$~\cite{palmer}), one sees from
the figure that $\etacp^\mu>8\%$ is accessible. To probe down to
$\etacp^\mu=2\%$ would require a 16-fold increase in the luminosity or
beam polarizations better than 75\%.

As we will discuss in the next section, meausurements of the electron EDM,
combined with well motivated assumptions about the slepton mass spectrum,
constrain the CP violation in the muon coupling to be very small within the
MSSM, usually
$<1\%$. However, the constraints on the $\tau$ coupling are much weaker.
So if the MSSM model is correct
a more theoretically promising,
though experimentally demanding, window for
viewing CP violation in the Higgs-lepton couplings is in the final state
$\tau$ polarization asymmetry from $h^0\to\tau^+\tau^-$.  (See
Ref.~\cite{soni} for a related discussion.)

Within the NMSSM (or other extensions of the MSSM Higgs
sector), much larger CP-violating effects are possible in the
Higgs-lepton couplings. Though the constraints on the mass spectrum and
CP-violating phases arising from the $e$EDM and $\mu$EDM,
reviewed in the following section, are essentially
identical to those in the MSSM, the
effective amount of CP violation in the Higgs-lepton couplings,
$\etacp^\ell$, is much larger for two reasons.  First, $\etacp^\ell$
in the NMSSM is unsuppressed by loop factors.  Second, we can
have a significant effect even for small $\tan\beta$. In the next
section, we will see that these properties permit much larger
values of $\etacp^\ell\sim 10\%$ to 100\% in the NMSSM.
Conservative muon collider design parameters are
already sufficient to see such large CP violation.

\section{Electric and magnetic dipole moments} \label{sec:edm}

The fundamental CP-violating phase of the $\mu$-term which is responsible for
the existence of a CP-violating Higgs-fermion coupling also contributes to the
electric dipole moments (EDMs) of the electron, muon and neutron.  In fact, the
diagrams of Fig.~\ref{fig:diagrams} contribute directly
to the EDM of an electron or muon when the external Higgs is
replaced by its vacuum expectation value
and a photon is attached to any charged line.
Because the experimental constraints on the muon and electron couplings are
so different, we will consider them each in turn.

\begin{figure}
\centering
\hspace*{0in}
\vspace{0in}
\centerline{
\epsfysize=2.5in
\hfill\epsfbox{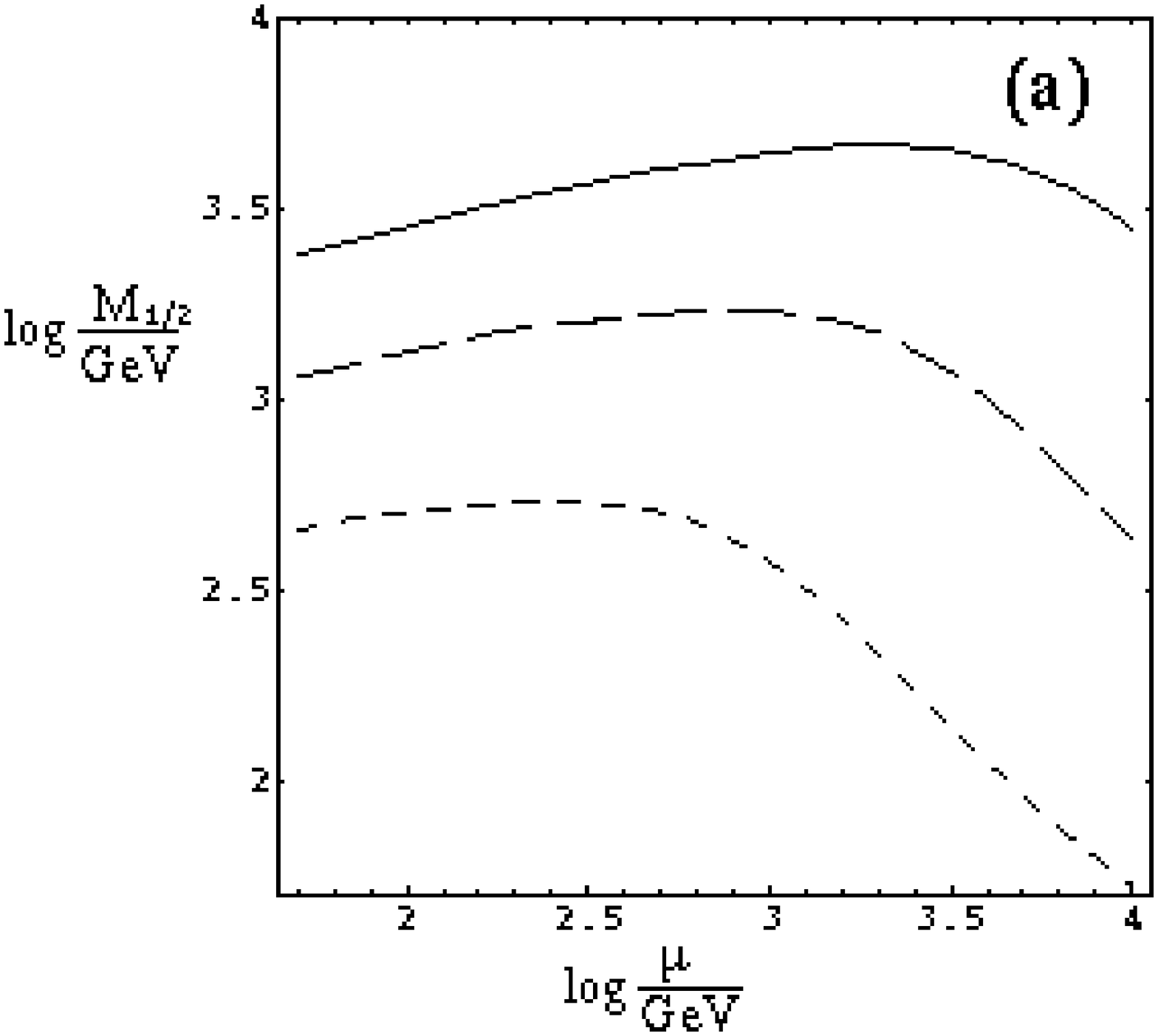}\hfill
\epsfysize=2.5in
\hfill\epsfbox{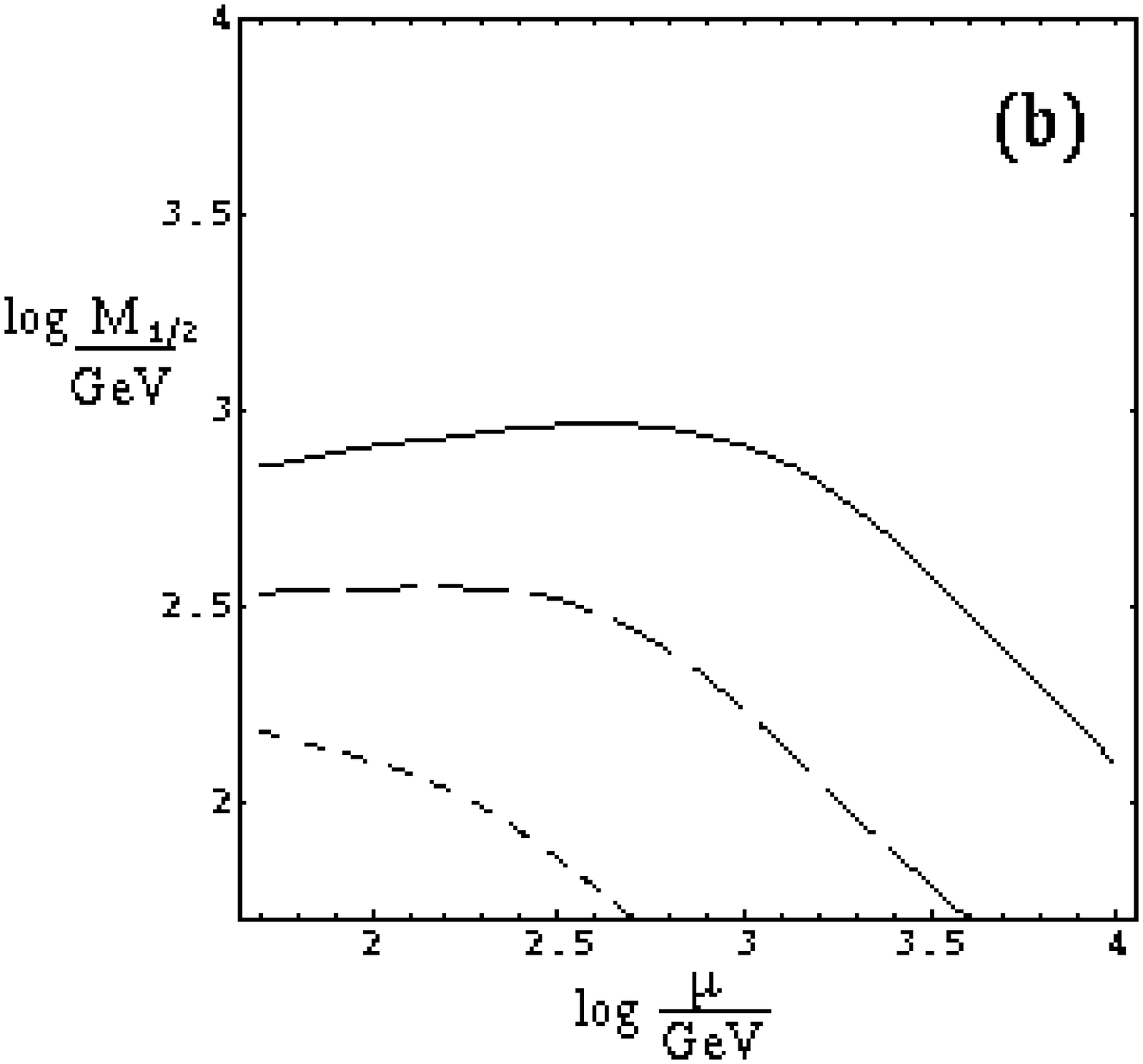}\hfill
}
\caption{EDM constraints for (a) $\tan\beta=50$ and (b)
$\tan\beta=2$, as a function of $|\mu|$ and the gaugino mass parameter,
$M_{1/2}$, both varying on a log scale from $50\gev$ to $10\tev$.  For
illustrative purposes, the soft scalar mass parameter is taken to satisfy
$M_0 = 2M_{1/2}$.  Solid, long-dashed, and short-dashed curves correspond
to $\arg(\mu) = \pi/4$, 0.1, 0.01 respectively.  Regions below the curves
are excluded by the current $e$EDM bound. }
\label{fig:edm}
\end{figure}

When $\ell=e$ in Figure~\ref{fig:diagrams}, a non-zero $e$EDM is generated.
Current experimental constraints on the $e$EDM are extremely
strong.  Specifically,
$d_e=(1.8\pm 1.6)\times 10^{-27}\,e\,$cm~\cite{commins}.  The effect of this
constraint on the SUSY parameter space is astonishing.  In
Figure~\ref{fig:edm}(a) we have shown the region of $\mu$ -- $M_{1/2}$
parameter space excluded (at 90\% C.L.) for $\tan\beta=50$
by the $e$EDM bound, following the calculation of Ref.~\cite{nath}\footnote{In
our evaluation of the l-loop contributions to the $\ell$EDMs
we have numerically 
diagonalized the full complex neutralino and slepton mass matrices.  No
mass-insertion approximation has been performed.}.   
Different contours correspond to differing sizes of
$\arg(\mu)$: $\pi/4$, 0.1, and 0.01.  Again, gaugino mass unification is
assumed, $M_i = \alpha_i M_{1/2}/\alpha_{\rm unif}$, and the soft
selectron mass parameter is taken to satisfy $M_0 = 2M_{1/2}$.
Figure~\ref{fig:edm}(b) shows the excluded regions for $\tan\beta =2$.

Plainly for $\CO(1)$ phases and $\tan\beta=50$, the masses of the SUSY
particles must be so heavy as to approach being unnatural. The end result,
in any case, is that one expects very little observable CP violation in
the coupling of the Higgs to electrons.  (Of course, simply
observing the CP-conserving Higgs-electron coupling is so challenging
that this is probably a moot point.)

For $\ell=\mu$, the situation is not as clear.
Current experimental bounds on the $\mu$EDM, $d_\mu=3.7\pm3.4\times
10^{-19}\,e\,$cm~\cite{RPP},
are not particularly constraining. Taken alone, the current bound on the
$\mu$EDM does little to limit the SUSY parameter space or
the size of $\etacp^\mu$ predicted in the MSSM.
Slightly more constraining is the muon magnetic dipole moment which is
already measured to an accuracy of approximately $10^{-8}$~\cite{RPP}.
It is well-known
that $g-2$ of the muon already excludes the MSSM with very large $\tan\beta$
and very light sleptons: for universal masses and $\tan\beta=50$, the sleptons
must lie above $250\gev$ at 90\% C.L.~\cite{moroi}.
In fact, it is the real part of the diagrams of
Fig.~\ref{fig:diagrams} that give such large contributions to $g-2$.

There are plans to improve both the $\mu$EDM measurement as well as that of
$(g-2)_\mu$ in the near future~\cite{e821}.
Over the next few years, direct limits on the $\mu$EDM should dramatically
improve, reaching a limit near $10^{-23}\,e\,$cm. Scaled by the appropriate
$m_e/m_\mu$, this limit approaches within a factor of 10 that already
obtained for the electron.
If a non-zero $\mu$EDM were to be measured, this would strongly bolster the
case for observable CP violation in the direct Higgs-muon coupling.
Conversely, non-observation would produce a strong bound, essentially
equivalent to the $\arg(\mu)=0.1$  $e$EDM bound shown in
Figure~\ref{fig:edm}(a), making it unlikely
that CP-violation in the MSSM Higgs-muon coupling could be observed.

We should take a moment to comment on the heavy mass limit.
Figure~\ref{fig:edm}(a) clearly shows that EDM constraints do/will require
either very small phases or very heavy sparticles. In the case of small
phases, $\etacp^\ell$ is similarly small. However, heavy masses
do not necessarily imply small $\etacp^\ell$, which is to say that
this is an example of a SUSY non-decoupling effect. It is easy to
understand why we do not find decoupling. In particular, if one takes
$m_A$ large along with the SUSY masses, leaving only a one Higgs doublet 
Standard Model at low energies, then these effects do decouple by
virtue of the fact that in a one Higgs doublet model all CP violation
can be eliminated through field redefinitions. The same is {\em not}
true in a two Higgs doublet version of the Standard Model, which is
what one has for light $m_A$. Now when the SUSY states are integrated 
out, they generate CP-violating $h^0\bar\ell\ell$ 
terms in the Lagrangian of the two
doublet model. Because this Higgs coupling is marginal
(renormalizable), those couplings are only logarithmically sensitive
to the SUSY scale. This is in contrast to the EDM operator,
$\bar\ell\sigma^{\mu\nu}\gamma^5\ell F_{\mu\nu}$,
which is irrelevant (non-renormalizable) and therefore flows quickly
to zero in the infrared as $1/M_{\rm SUSY}$.

Do such large SUSY masses make sense?
Perhaps. The main ``esthetic'' or ``naturality'' constraint on supersymmetric
partner masses is that they not induce too large a correction to the
Higgs (mass)$^2$ parameter through diagrams which grow quadratically
with the mass of the superpartner.  The one-loop contributions of the
selectrons and smuons
to the Higgs (mass)$^2$ parameter are proportional
to their relatively small Yukawa couplings,
allowing one to satisfy the naturality constraint
with masses as large as $1000\tev$.   (By way of contrast,
the third generation squarks and sleptons as well as the
gauginos, because of their $\order(1)$ couplings to the Higgs,
must have masses  $\lsim 1\tev$.)  However, there are two-loop contributions
involving the light generation sfermions not proportional to their small
Yukawa couplings, and naturality~\cite{dimop-nelson} and vacuum
stability\cite{murayama}
require that these sfermion masses be below about $(5-20)\tev$.
Thus it is not immediately clear whether one should consider smuon
masses such as those demanded by Figure~\ref{fig:edm}(a) to be
unnatural or not.  (Note that if the gaugino masses are $\ltap 1\tev$,
but 1st and 2nd generation scalars have masses in the $5\sim10\tev$ range,
then $\etacp^\ell$ in the MSSM is suppressed by a factor of
$(m_2/m_{\tilde \mu})\sim 1/10$; see Eqs.~(\ref{eq:A1}) -- (\ref{eq:A3}).)

However, as we now explain, under some reasonable assumptions
concerning the SUSY spectrum, the constraint on the $e$EDM can be used
to constrain the $\mu$EDM already.  This is because the phase that enters
the leptonic coupling via $\arg(\mu)$ is universal -- all sleptons receive
exactly the same phase. There can be non-universal phases coming from the
trilinear $A$-terms, but they are not enhanced at large $\tan\beta$, and
unless one assumes some kind of cancellation, the
size of the CP-violating phases of the various sleptons are correlated.
Furthermore, if the smuon and selectron are approximately
degenerate, then the constraint
on the $e$EDM translates directly into a constraint on the muon-Higgs
coupling.  It implies for one thing that $d_\mu<10^{-24}\,e\,$cm, which means
that the BNL E821 experiment~\cite{e821} looking for a non-zero
$\mu$EDM should obtain a null result.  It also means that
the current limit on the $e$EDM already constrains the CP violation in the
MSSM Higgs-muon coupling to be unobservable at a Higgs factory, even without
any improvement in the $\mu$EDM measurement.

Why should we assume that the smuons and sleptons will be degenerate?
Non-degenerate sleptons generically lead to large flavor-changing neutral
currents (FCNC's),
specifically $\mu\to e\gamma$.  Very heavy
sleptons could also account for the lack of FCNC's in leptonic processes, but
if the sleptons are non-degenerate then it is natural to expect the same for 
the squarks, and there the constraints are much stronger. In particular, in the
presence of generic CP-violating phases, $d$-squark masses would have to
exceed approximately $5000\tev$~\cite{kkbar} in order to agree with the
measured CP violation in the kaon system, specifically  
$\eps_K\simeq2\times10^{-3}$. This is far above any possible
definition of a natural squark mass and we do not consider this possibility
further. There is also the possibility that the new contributions to $\eps_K$
could be eliminated through alignment of the quark/squark mass
matrices~\cite{nir},
but to our knowledge no very attractive model has been built along
these lines. Thus we are left with degeneracy in the squark
sector as by far the most attractive solution to the dual problems of FCNC's
and $\eps_K$, and by extension degeneracy becomes the most attractive
scenario for the sleptons as well.

The preceding discussion of degeneracy really only applies to the first
two generations of sparticles. The third generation, thanks to its small
quark mixings with the other generations, has suppressed contributions
to FCNC's and $\eps_K$. It would not be difficult to imagine, for example, that
the first two generations of sleptons are degenerate and heavy, while the
third is much lighter. In that case the constraint from the $e$EDM would not 
limit CP violation in the $\tau$ sector and so the decay $h^0\to\tau^+\tau^-$
may be a likely place to observe violation of CP. (If $m_{\tilde\tau}=
m_{\tilde e}$, then the $e$EDM constraint does apply, leaving little room
for CP violation in the Higgs-$\tau$ coupling.)

The situation for the NMSSM and other extensions is slightly different.
First, we emphasize that the EDM {\em constraints}\/  on the (tree-level)
phase are essentially identical (for a given SUSY spectrum) to
the constraint on the size of CP-violating phases in the
MSSM.  Indeed, after suitable field
redefinitions, this phase of the NMSSM can be moved into a phase
of the effective $\mu$-term, $\lambda\vev{N}$.  Consequently,
{\em precisely} the same diagrams contribute to the EDMs
in the NMSSM as in the MSSM.  The only difference, a minor one,
is that the neutralino mass matrix is now a $5\times5$ matrix;
in the limit of fixed $\mu\equiv \lambda\vev{N}$ and $k\vev{N}
\gg \lambda v$, the contributions involving the $N$ component of the
neutralino to the EDM are suppressed, and one recovers exactly the
structure arising in the MSSM.

Nevertheless, as we discussed in Sections~\ref{sec:nmssm}-\ref{sec:collider},
there are two important differences between the MSSM and NMSSM.
First, in the NMSSM the CP violation in the Higgs-lepton couplings is
neither enhanced nor diminished by $\tan\beta$ since it is a tree-level
effect. Therefore one does not require very large values of $\tan\beta$ in
order to obtain observable violations of CP. From the point of view of the
EDM constraints, however, it is advantageous to have small $\tan\beta$.
This is clear when comparing Figures~\ref{fig:edm}(a) and (b). For the
same phase, the mass bounds at $\tan\beta=2$ are roughly four times
weaker than those at $\tan\beta=50$. Thus the constraint coming
from naturalness (\ie, demanding light scalar masses) is less restricting
at small $\tan\beta$, allowing in turn larger underlying phases and thus
larger $\etacp$.

A second difference is that 
in the NMSSM CP-violating couplings of the Higgs particles to
fermions occur at tree-level. Thus there are no loop suppressions which
suppress $\etacp$ with respect to the EDM. 

The final effect of these differences is that for scalar masses in the
range 1 to $3\tev$, the underlying CP-violating phases in the NMSSM can be 
$\CO(1)$ and
thus $\etacp^\ell\sim\CO(1)$ as well. Even for sleptons in the $500\gev$
mass range, it is quite natural to expect $\etacp^\ell\sim10\%$, which
can easily be probed in a single year at a muon collider of
current conservative design parameters (see Figure~\ref{fig:ALR}).
This is contrasted with the much smaller
value ($\etacp^\ell\sim1\%$) expected in the MSSM.
Indeed, one can take advantage of this
difference and use a combination of EDM and CP-violating
muon-Higgs coupling measurements to discriminate between the MSSM
and its extensions.


\section{Conclusions}

In this paper we have shown that at 1-loop a
potentially large CP-violating coupling of the
Higgs to SM fermions is induced in the MSSM.  The CP-violating
coupling to muons, $\etacp^\mu$, could be accessed cleanly
through the polarization-dependent
production asymmetry at a muon collider operating on the Higgs
resonance.  However, by imposing reasonable theoretical expectations,
the motivations for which were discussed in Section~\ref{sec:edm}, 
together with the current bounds on the $e$EDM, we  found
stringent constraints on the size
of $\etacp^\mu$ in the MSSM.  We argued that the CP-violating coupling
of third generation fermions to the Higgs could be substantial nevertheless.
In simple, natural extensions of the MSSM, such as
the NMSSM, CP-violating couplings of $h^0$ to SM fermions occur at
tree level, and large CP violation is plausible for the Higgs couplings of
all three generations of SM fermions, even after imposing the
$e$EDM constraints.
In particular, CP-violating signals 
at a muon collider of $\CO(100\%)$ are not ruled out.

\section*{Acknowledgments}

This research was
supported in part by U.S. Department of Energy contract
\#DE-FG02-90ER40542, and by the W.M.~Keck Foundation. CK wishes to
acknowledge the generosity of Helen and Martin Chooljian. JMR is supported
in part by an Alfred P.~Sloan Foundation fellowship.
We would like to thank T.~Ibrahim, H.~Murayama, M.~Peskin, and
N.~Polonsky for helpful conversations.


\begin{thebibliography}{99}

\bibitem{maekawa}
        N.~Maekawa, \PLB{282}{92}{387}.

\bibitem{banks}
        T.~Banks, \NPB{303}{88}{172};\\
        R.~Hempfling, \PRD{49}{94}{6168}.

\bibitem{hrs}
        L.~Hall, R.~Rattazzi and U.~Sarid, \PRD{50}{94}{7048}.

\bibitem{HHG}
        See, \eg, J.~Gunion \etal,
        {\em The Higgs Hunter's Guide}, Addison-Wesley, 1990.

\bibitem{fayet}
        P.~Fayet, \NPB{90}{75}{104}.

\bibitem{haba}
        N.~Haba, Prog.\ Theor.\ Phys.\ {\bf 97} (1997) 301.

\bibitem{gunion}
        B.~Grzadkowski and J.~Gunion, \PLB{294}{92}{361}.

\bibitem{soni}
        D.~Atwood and A.~Soni, \PRD{52}{95}{6271}; \\
        A.~Pilaftsis, \PRL{77}{96}{4996}; \\
        B.~Kamal, W.~Marciano and Z.~Parsa, {\tt hep-ph/9712270}.

\bibitem{palmer}
        R.~Palmer (for the Muon Collider Collaboration), {\tt physics/9802005}.

\bibitem{commins}
        E.~Commins \etal, Phys.\ Rev.\ {\bf A50} (1994) 2960.

\bibitem{nath}
        T.~Ibrahim and P.~Nath, \PRD{57}{98}{478}. \\
        See also Y.~Kizukuri and N.~Oshimo, \PRD{46}{92}{3025}.

\bibitem{RPP}
        R.~Barnett \etal (Particle Data Group), \PRD{54}{96}{1}.

\bibitem{moroi}
        T.~Moroi, \PRD{53}{96}{6565};\\
        M.~Carena, G.~Giudice, and C.~Wagner, \PLB{390}{97}{234}.

\bibitem{e821}
        Y.~Semertzidis \etal (E821 Collaboration), {\it AGS Expression
        of Interest: Search for an Electric Dipole Moment of the Muon},
        Sept.\ 1996.  See also {\tt http://www.phy.bnl.gov/g2muon/home.html}.

\bibitem{dimop-nelson}
        S.~Dimopoulos and G.~Giudice, \PLB{357}{95}{573}; \\
        A.~Cohen, D.~Kaplan and A.~Nelson, \PLB{388}{96}{588}.

\bibitem{murayama}
        N.~Arkani-Hamed and H.~Murayama, \PRD{56}{97}{6733}.

\bibitem{kkbar}
        F.~Gabbiani \etal, 
        \NPB{477}{96}{321}; \\
        J.~Bagger, K.~Matchev and R.~Zhang, \PLB{412}{97}{77}.

\bibitem{nir}
        Y.~Nir and N.~Seiberg, \PLB{309}{93}{337}; \\
        Y.~Nir and R.~Rattazzi, \PLB{382}{96}{363}.

\end{thebibliography}
\end{document}